\documentclass{ws-procs9x6}

\begin{document}

\title{DVCS at HERMES: Recent Results}

\author{F. Ellinghaus\footnote{\uppercase {T}his work is
supported in part by the \uppercase {US}  \uppercase {D}epartment of 
 \uppercase {E}nergy.}}

\address{University of Colorado, \\
Department of Physics,   \\
Boulder, Colorado 80309-0390, USA \\ 
E-mail: Frank.Ellinghaus@desy.de}

\author{for the HERMES Collaboration}

\begin{abstract}
Hard exclusive reactions are the tool to learn about generalized parton
distributions, which provide a more complete parametrization of the nucleon than
the ordinary parton distribution functions. 
Recent measurements by the HERMES collaboration
of the exclusive production
of photons, i.e., Deeply-Virtual Compton Scattering, 
are summarized and compared to model calculations, 
 focusing on the measurements and model comparisons relevant to the extraction of quark
orbital angular momentum and on the measurements on heavy nuclei.
\end{abstract}


\bodymatter

\section{Introduction}
Similar to the case of inclusive and semi-inclusive DIS, where the 
nucleon structure is described using {\bf P}arton {\bf D}istribution 
{\bf F}unctions (PDFs), hard exclusive reactions can be expressed
in terms of {\bf G}eneralized {\bf P}arton {\bf D}istributions
(GPDs)\cite{Mue94,Ji97a,Rad97}.
The PDFs and elastic nucleon {\bf F}orm {\bf F}actors (FFs) are included in the GPDs 
as the limiting cases and moments of GPDs, respectively\cite {Ji97a}.
While FFs derived in elastic
scattering describe the transverse location of partons inside the nucleon and 
PDFs describe their longitudinal momentum distribution, GPDs 
are able to provide information on both at the same time. Thus exclusive
reactions are able to give a certain 3--dimensional picture of the nucleon
structure~\cite{Bur00,Bel02b,Ral02}.
In particular, GPDs offer for the first time
a possibility to determine the total angular momentum 
carried by the quarks in the nucleon~\cite{Ji97a}.

Below recent HERMES measurements on the 
hard exclusive electroproduction of real photons
({\bf D}eeply--{\bf V}irtual {\bf C}ompton {\bf S}cattering, DVCS) are
summarized and compared to model calculations.
The data has been taken with polarized and unpolarized gas targets
using the HERMES spectrometer \cite {Ack98}
at the HERA electron/positron--proton collider at DESY, which offers
longitudinally polarized 27.6 GeV electron and positron beams.

\section{GPD $H$ via beam-charge and beam-spin asymmetries}
DVCS amplitudes can be measured through the interference
between the DVCS and {\bf B}ethe--{\bf H}eitler (BH) processes, 
in which the photon is radiated from a parton in the former and
from the electron in the latter process. 
Both processes have an identical final state, 
i.e., they are indistinguishable, and thus 
give rise to an interference term $I$.
The photon production cross section 
depends on the Bjorken scaling variable $x_B$, 
the squared virtual--photon four--momentum $-Q^2$,
the squared four-momentum transfer $t$ or the reduced four-momentum 
transfer $t^\prime = (t-t_{min})$ to the target, and
the azimuthal angle $\phi$ defined as the angle between the lepton
scattering plane
and the (virtual and real) photon production plane. 
For an unpolarized proton target, and at leading twist, the interference term
is given by \cite{Die97}
\begin{eqnarray} 
\label {I}
I \propto -C \, 
[ \, a \, \cos \phi \, \mathrm{Re} {\cal M}^{1,1}
- b \, P_l \, \sin \phi  \, \mathrm{Im} {\cal M}^{1,1} 
\, ], 
\end{eqnarray}  
where 
the lepton beam has longitudinal polarization $P_l$
and charge $C = \pm 1$,
and $a$ and $b$ are functions of the ratio of the longitudinal to transverse 
virtual--photon flux. 
The real (imaginary) part of the DVCS amplitude $M^{1,1}$ can be accessed 
by measuring the $\cos \phi$ ($\sin \phi$) dependence of a cross section asymmetry with
respect to the charge (spin) of the lepton beam.
At HERMES kinematics, the DVCS amplitude $M^{1,1}$ gives access to the GPD~$H$.
Details and full equations are given in Ref.~\refcite {Bel02a}.

The event selection at HERMES requires events with
exactly one photon and one charged track, 
identified as the scattered lepton, with $Q^2 >$1~GeV$^2$. 
For data taken prior to 2006
the recoiling proton is not detected 
and exclusive events are identified by the fact that the  
missing mass $M_x$ of the reaction $e p \rightarrow e \gamma X$ corresponds to the proton
mass. 
Due to the finite energy resolution
the exclusive sample is defined as $-1.5 < M_x < 1.7$~GeV. 
A recoil detector used during the recent data taking should reduce the underlying background
from presently about 15\% to less than 1\%~\cite{Roberto}.

\begin{figure}[t]
\includegraphics[width=0.53\textwidth]{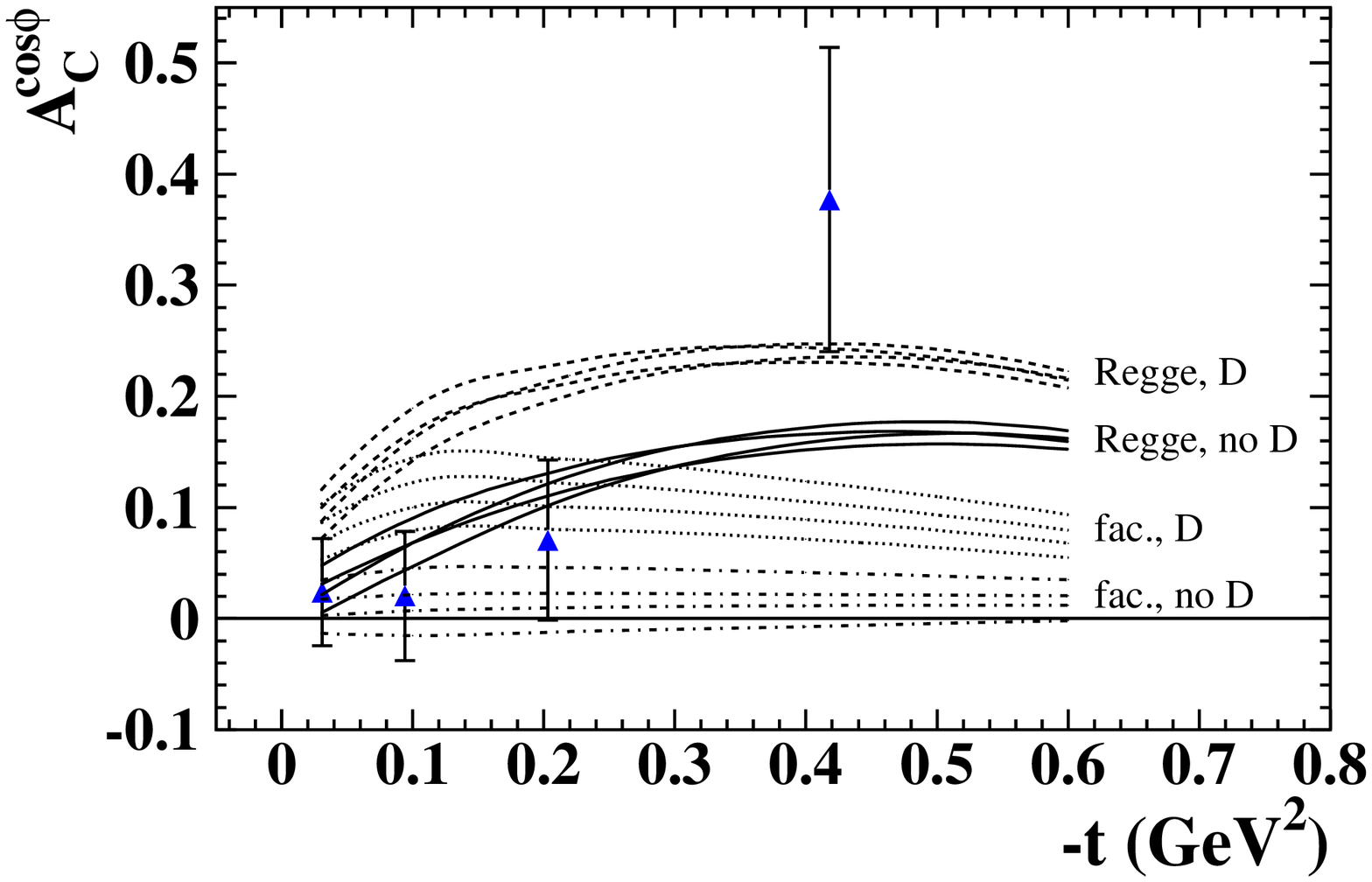}
\includegraphics[width=0.46\textwidth]{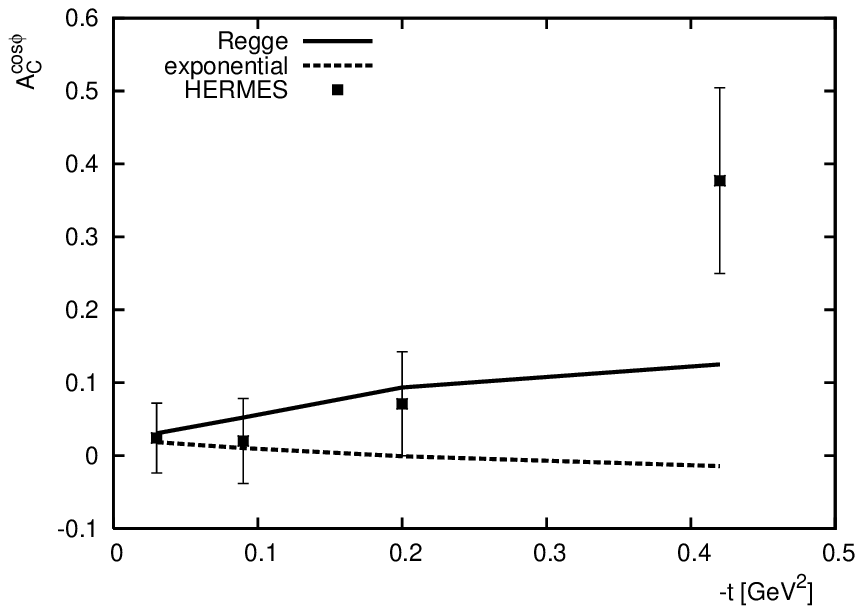}
\caption [*]{
The $\cos \phi$ amplitude of the beam--charge asymmetry \cite{Air06} on
hydrogen
as a function of $-t$. The error bars show the statistical and systematic
uncertainties added in quadrature. 
The calculations in the left panel are based on a double-distribution GPD model using a factorized (fac.)
 or a Regge--inspired (Regge) $t$--dependence with (D)
    or without (no D) a D-term contribution. The parameters $b_v$ and $b_s$ are each set to either unity
   or infinity.
Using the resulting 16 sets of model parameters, the calculated asymmetries fall into four
   main groups. The variations within these groups are due to the different
 settings for the $b$ parameters.
The calculations in the right panel are based on a dual-parametrization GPD model~\cite{Guz06}.}
\label{bca_t}
\end{figure}

\begin{figure}[t]
\includegraphics[width=0.495\textwidth]{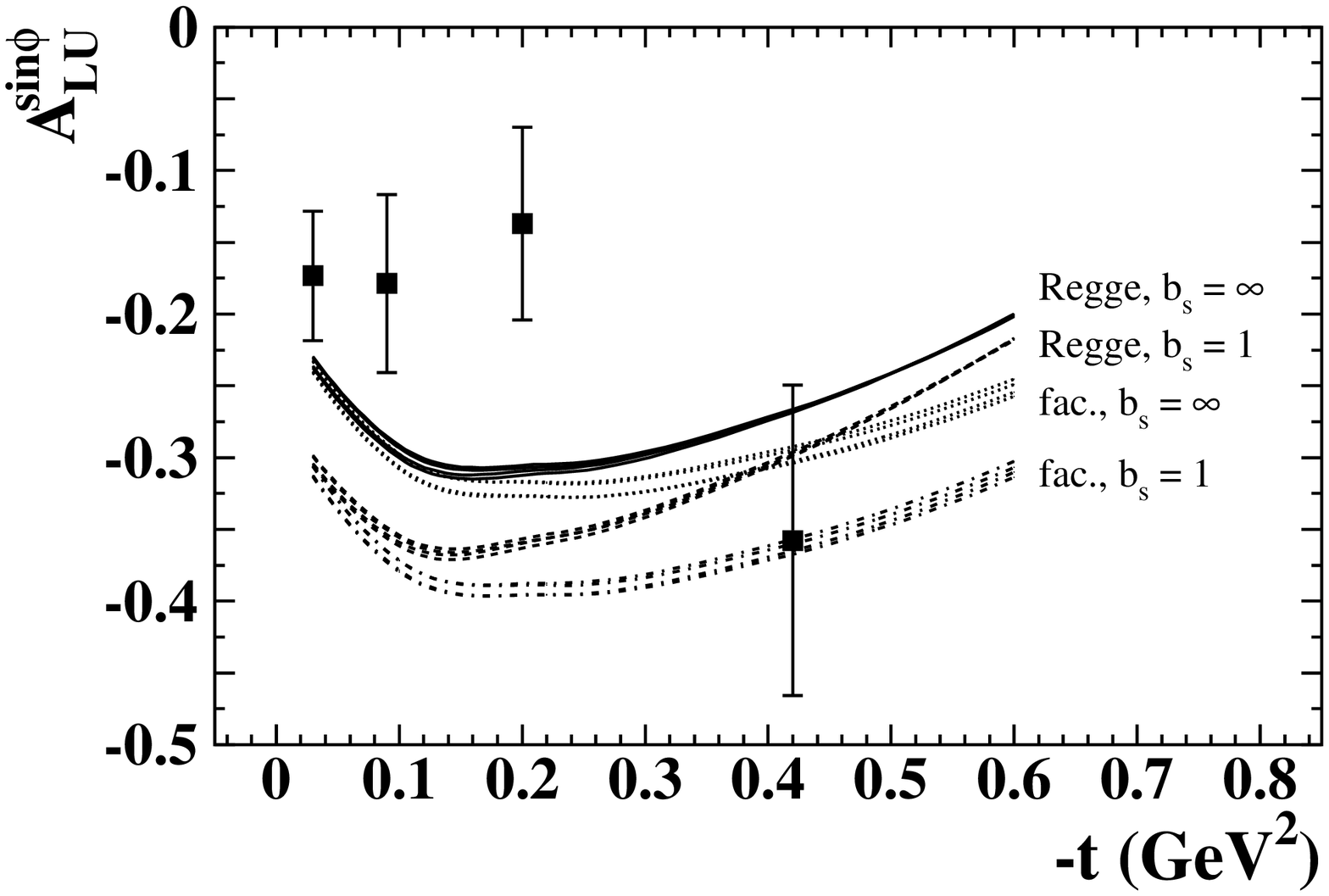}
\includegraphics[width=0.495\textwidth]{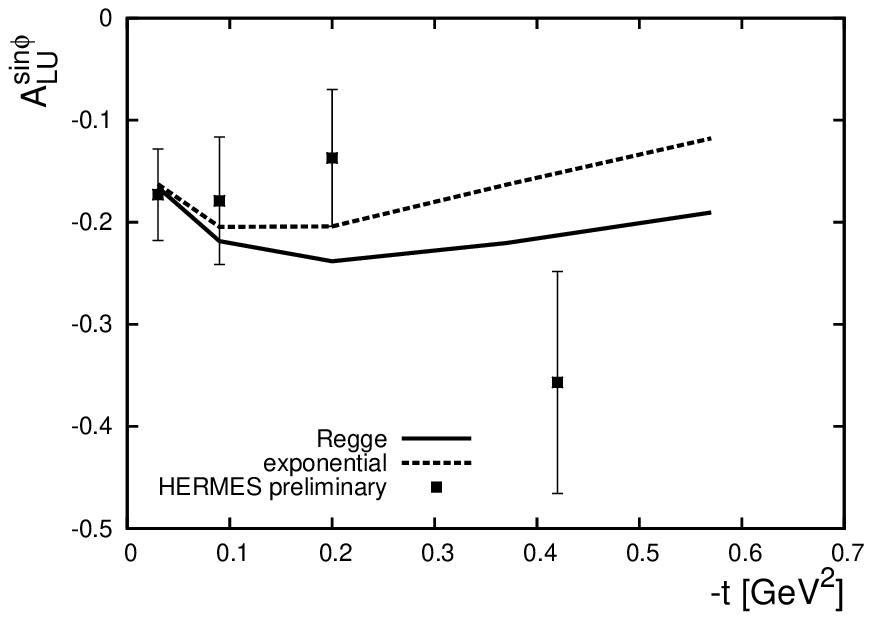}
\includegraphics[width=0.495\textwidth]{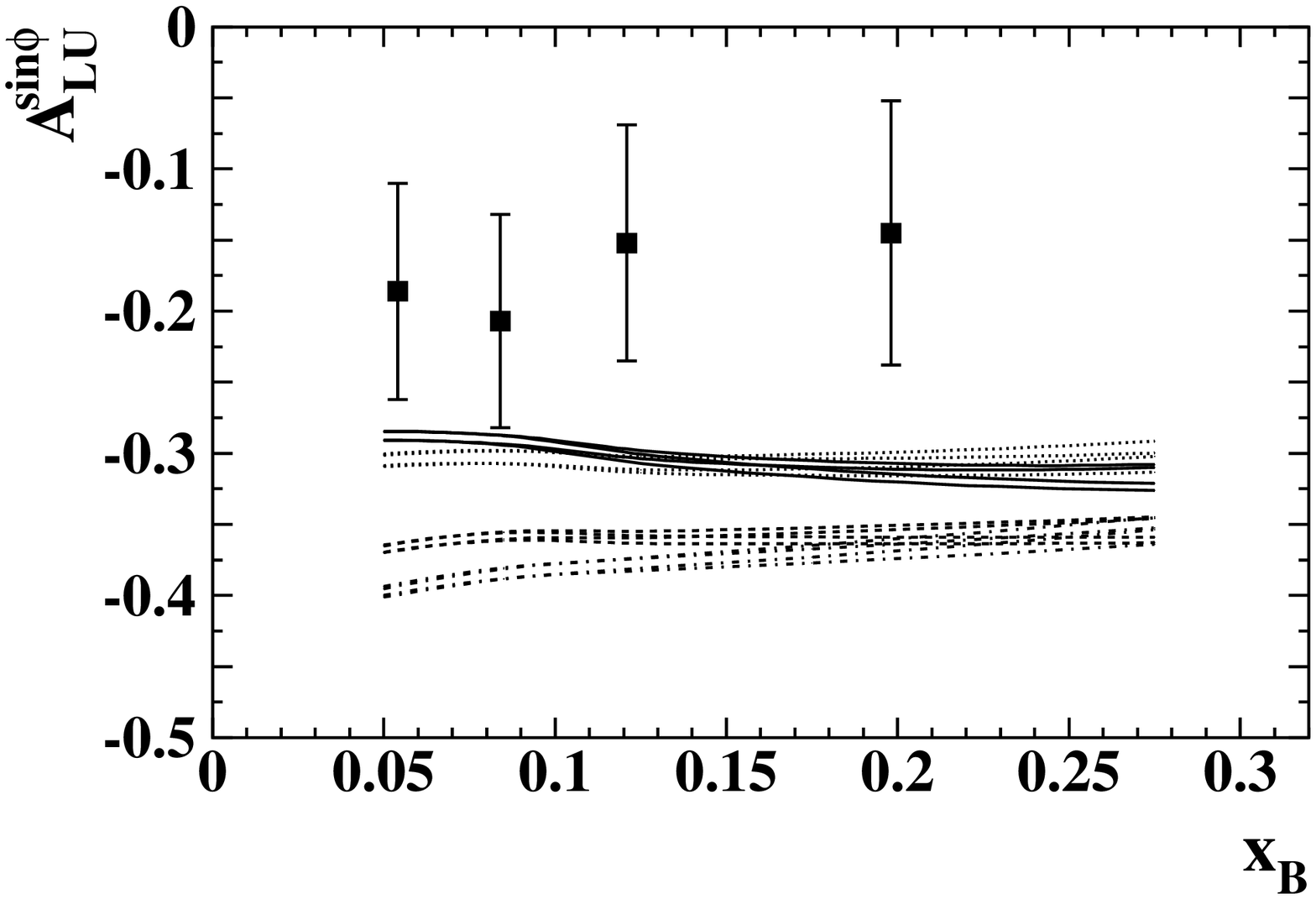}
\includegraphics[width=0.495\textwidth]{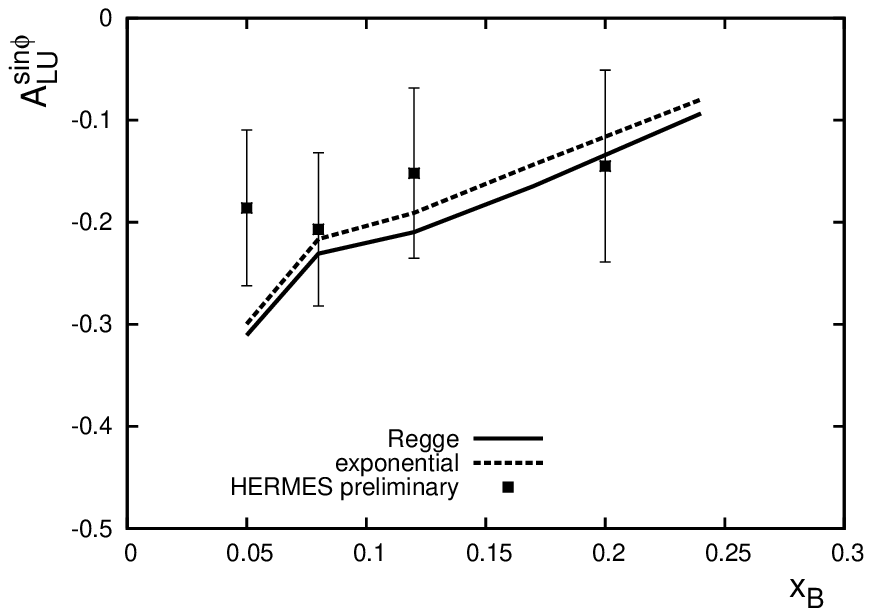}
\includegraphics[width=0.495\textwidth]{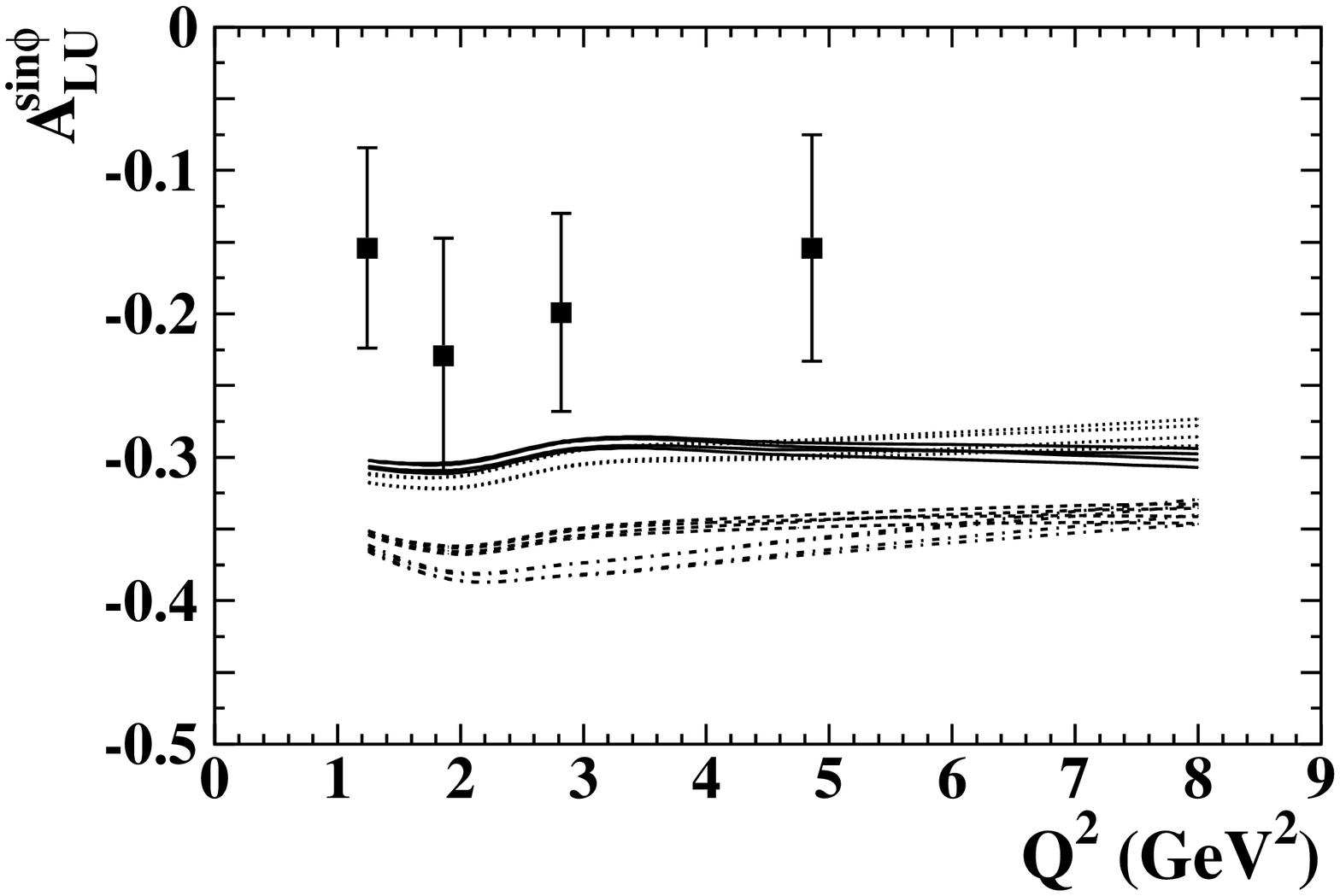}
\includegraphics[width=0.495\textwidth]{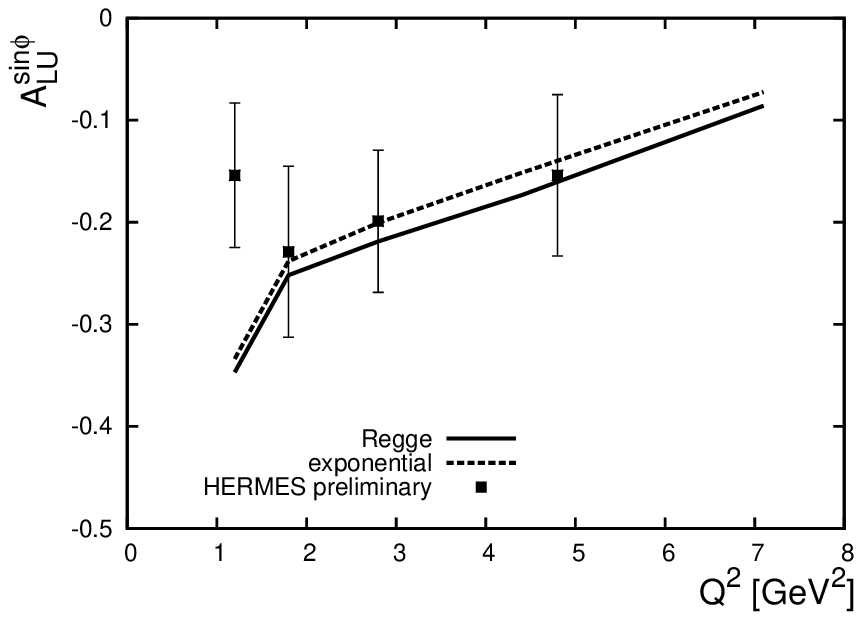}
\caption [*]{
The $\sin \phi$ amplitude of the beam--spin asymmetry on hydrogen from the 1996-2000 data
   as a function of $-t$, $x_B$ and $Q^2$. 
The error bars show the statistical and systematic
uncertainties added in quadrature.
The calculations in the left panel are based on a double-distribution GPD model using a factorized (fac.)
 or a Regge--inspired (Regge) $t$--dependence with
    or without a D-term contribution. The parameters $b_v$ and $b_s$ are each set to either unity
   or infinity.
Using the resulting 16 sets of model parameters, the calculated asymmetries fall into four
   main groups, whereby the BSA appears to be insensitive to the D-Term and to the
   value of $b_v$. 
The calculations in the right column are based on a dual-parametrization GPD model~\cite{Guz06}.}
 \label{bsa_kindep_vgg_guzey}
\end{figure}

The beam--spin asymmetry (BSA) and the beam--charge asymmetry (BCA) as a
function of $\phi$ are calculated as
\begin{equation} 
A_{LU} (\phi) = \frac {1}{< \left | P_l \right | >} \, 
\frac {\overrightarrow N (\phi) - \overleftarrow N (\phi)}
{\overrightarrow N (\phi) + \overleftarrow N (\phi)}, \qquad
A_C(\phi) = \frac{ N^+(\phi) - N^-(\phi)}
{N^+(\phi) + N^-(\phi) },
\label{bsabca}
\end{equation}
with the normalized yields $\overrightarrow N$ ($\overleftarrow N$) or $N^+$ ($N^-$)
using a beam with positive (negative) helicity or a positron (electron) beam, respectively.
The BSA (BCA) on the proton as a function of $\phi$ has been extracted at
HERMES \cite{Air01} (\cite{Air06}),
whereby
the predominant $\sin \phi$ ($\cos \phi$) dependence expected from
Eqn.~\ref{I} has been observed.
The $\cos \phi$ amplitudes of the BCA on 
hydrogen as a function of $-t$ derived from a fit to the BCA in each $-t$
bin~\cite {Air06} are shown in Fig.\ref{bca_t}, and the 
recent preliminary BSA result on the kinematic dependences of the $\sin \phi$ amplitudes is 
shown in Fig.\ref{bsa_kindep_vgg_guzey}.

The model calculations shown in the left panel (left column) of
Fig.\ref{bca_t} (Fig.\ref{bsa_kindep_vgg_guzey})
are based on a double-distribution GPD model 
described in Refs. \refcite {Van99,Goe01}. 
Since these data are exclusively sensitive to the GPD $H$ as mentioned above,
the theoretical calculations shown 
were derived by only varying the model parameters for the GPD $H$ 
in the underlying code \cite {Van01}
in order to calculate the asymmetries at the average kinematics of every bin.
The model parameters differ with respect to including or neglecting the so--called
D--term \cite {Pol99} in the GPD model and whether the $t$--dependence of the GPD $H$ is 
calculated in either the simplest ansatz where 
the $t$--dependence factorizes from the $t$--independent part
or in the Regge--motivated ansatz. In addition,
the so-called skewness parameters $b_{v}$ and $b_{s}$ in the profile function~\cite{Mus00}
have been set to either unity or infinity, the latter value corresponds to a
skewness independent ansatz for the GPD.
The BCA data appears to disfavor the four parameters sets with the Regge-inspired $t$-dependence
and the D--term contribution as well as the factorized one with the D-term which gives the
largest asymmetry ($b_v=1$, $b_s=\infty$). For the BSA
all models seem to overshoot 
the absolute size of the $\sin \phi$ amplitude. 
The calculations based on the dual parametrization GPD
model~\cite{Guz06} shown in the right panel (right column) of Fig.\ref{bca_t} 
(Fig.\ref{bsa_kindep_vgg_guzey}) are all in rather good agreement with the data.

\section {GPDs $H$ and $E$ via transverse target-spin asymmetry}
Within the next few years HERMES will be able to provide sufficient data to largely constrain
the GPD $H$ in the kinematic region of the experiment. It is natural to ask to what extent the GPD $E$
can be accessed, which is the other important GPD necessary in order to determine $J_q$,
the total orbital angular momentum of quarks in the nucleon~\cite{Ji97a}. 
For an unpolarized proton target the contribution from the GPD $E$ is suppressed with respect to $H$, 
but this is different for transverse target polarization \cite{Bel02a,Die03}.
Using an unpolarized beam (U) and a transversely (T) polarized target, 
a $\sin{(\phi-\phi_S)}\cos{\phi}$ modulation in the DVCS
transverse target-spin asymmetry (TTSA) gives access to a combination of the
GPDs $H$ and $E$ \cite {Ell05}. Here $\phi_s$ denotes the azimuthal angle of the target polarization vector with
respect to the lepton scattering plane.

\begin{figure} [t]
\includegraphics[width=0.9\textwidth]{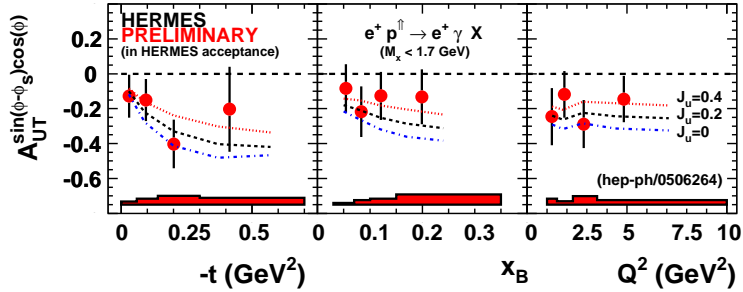}
 \caption{The $\sin (\phi-\phi_S) \cos \phi$ amplitude of the transverse target-spin asymmetry 
as a function of $-t$, $x_B$ and $Q^2$ in
          comparison to theoretical predictions from Ref.~\refcite{Ell05}.}
 \label{trans_proj}
\end{figure}
\begin{figure}[t]
\begin{center}
\includegraphics[width=0.7\textwidth]{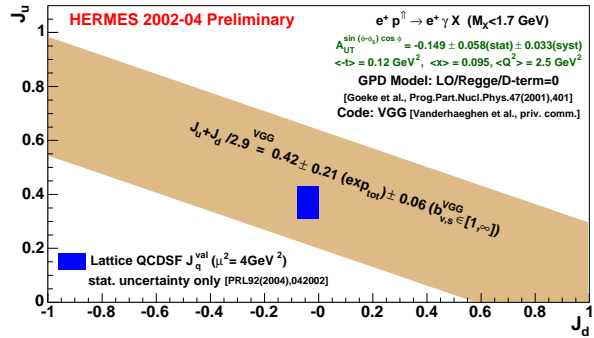}
\caption {Model dependent constraint on the quark orbital angular
  momenta
$J_u$ and $J_d$.}
 \label{jujd}
\end{center}
\end{figure}

The results from HERMES data collected on a transversely polarized hydrogen
target
are shown in Fig.~\ref {trans_proj}.
They agree with the model calculations~\cite{Ell05} shown in the same figure,
which have been calculated for various values of $J_u$. 
Based on $u$-quark dominance the $d$-quark total angular momentum has been assumed to be zero.
Since it was realized that the model calculations 
are largely insensitive to all model parameters but $J_u$ and $J_d$, it is
possible to constrain $J_u$ and $J_d$ even though other model parameters are 
largely unconstrained~\cite{Ell05}. 
For example, the model calculations 
in Fig.~\ref{trans_proj}
show little variation if calculated
for any of the 16 parameter sets which give very different values
for the BCA as shown in Fig.\ref{bca_t}. 
Hence the four parameter sets with the  Regge--inspired $t$--dependence and without a
contribution of the D--term (solid lines in the left panel of Fig.\ref{bca_t}),
have been chosen in order to determine the parameter space allowed for $J_u$ and $J_d$. 
For different values of $J_u$ and $J_d$ these sets have been compared to the
DVCS TTSA data shown in Fig.~\ref{trans_proj},
leading to a first model dependent constraint on $J_u$ versus $J_d$ shown in Fig.\ref{jujd}~\cite{Ye06}.
Further improvement can be expected, taking into account that the GPD $H$ and
therefore the available theoretical
models will be well constrained by the upcoming HERMES 
data, and that the data shown here is less than half of the
data taken on the transversely polarized hydrogen target.
Due to the good agreement between the calculations in the dual-parametrization
model and the BSA and BCA results as shown above, the next natural step is to
also calculate the allowed parameter space for $J_u$ and $J_d$ within this
model. This effort is presently ongoing, 
but calculations for various values of $J_u$ at a fixed $J_d = 0$~\cite{Guz06}
already indicate that the dual model favors smaller values for $J_u$
than the double-distribution one.

\section{DVCS on Nuclei}
\begin{figure}[t]
\includegraphics[width=0.595\textwidth]{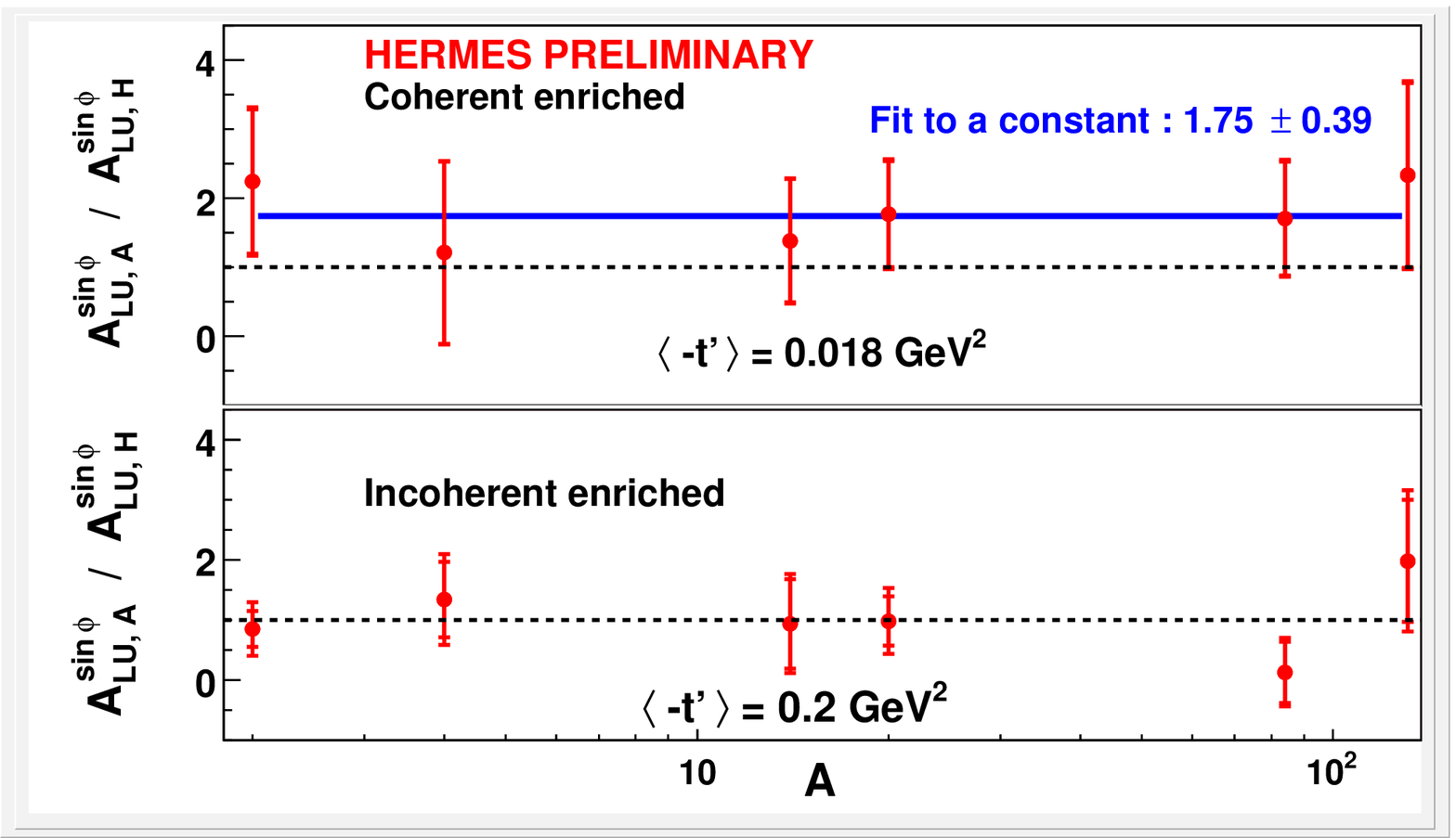}
\includegraphics[width=0.395\textwidth]{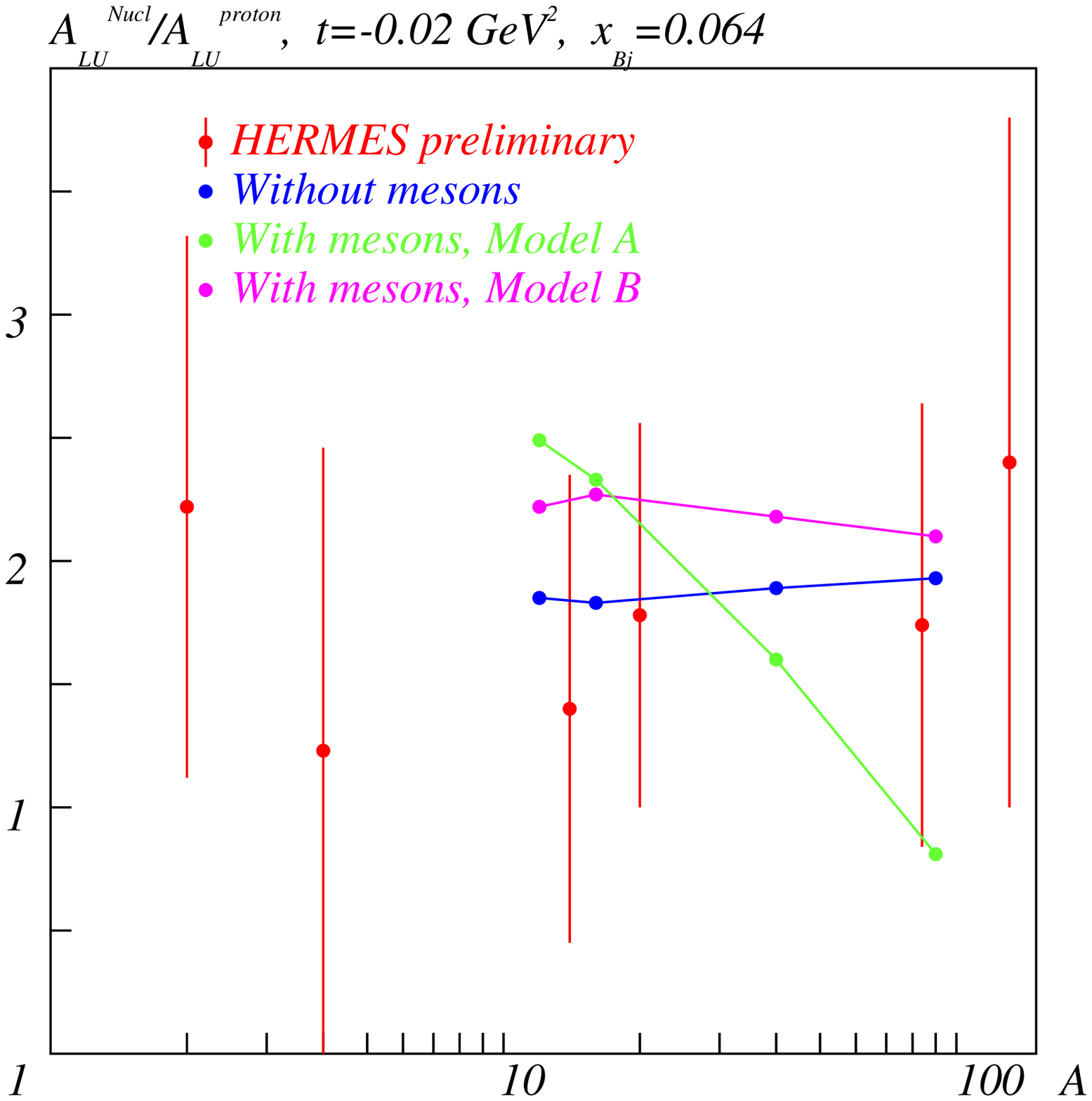}
\caption [*]{The $\sin \phi$ amplitude of the BSA on D, He, N, Ne, Kr, Xe divided by
  the one on the proton in two different kinematic regions dominated by either
  coherent or incoherent processes (left panel).
The 'coherent enriched' result (left panel, upper row) is compared to model 
calculations~\cite{Guz06b} in the right panel.}
\label{nuclei}
\end{figure}

Beam-Spin Asymmetries have also been measured on deuterium and various heavier
nuclei (He, N, Ne, Kr, Xe). Coherent processes are dominant for small values
of $-t$, while at larger values incoherent scattering on the individual protons
and neutrons dominates. Based on MC studies a region of small $-t^\prime$ (``coherent
enriched'')
has been selected for each nucleus in order to have similar average kinematic
values for all targets ($\langle -t^\prime \rangle =0.018$ GeV$^2$, $\langle Q^2
\rangle \approx 1.7$ GeV$^2$, $\langle x_B \rangle \approx 0.065$).
According to the MC studies the fraction of coherent processes is
approximately 82\% for all but the lighter targets (D, He). 
Similarly, an ``incoherent enriched'' sample has
been selected with an average $-t^\prime$ of 0.2~GeV$^2$. The ratio of the BSA on the nuclei
divided by the one on the proton at the same average kinematics is shown in
the left panel of Fig.\ref{nuclei}. A simple fit to a constant for the coherent enriched
sample yields a value above unity by two sigma, while the incoherent enriched
sample shows asymmetries very similar to the one on the proton. Note that this is in
agreement with earlier HERMES preliminary results~\cite{Ell02c} where a value consistent
with unity was found when comparing the BSA on deuterium and neon to the one
on the proton in the full $t$ range.
The BSAs on nuclei are expected to be similar to the one on the proton in the 
incoherent enriched sample since scattering on the protons inside the nuclei
should dominate due to the fact that the BH process on the neutron is
suppressed.
The fit value found for the coherent enriched sample is consistent with a
very basic prediction of 5/3 for Spin-0 and Spin-1/2 targets, based on the
ratio of the involved valence-quark charges squared~\cite{Kir03}, as well
as with calculations done specifically for neon and krypton~\cite{Guz03}.
Also a
prediction of R= 1-1.1 for Helium~\cite{Liu05} is in agreement with the
measurement. The result of a recent calculation~\cite{Guz06b} is shown 
in the right panel of Fig.\ref{nuclei}, whereby one of the three models is disfavored by the data.

\section{Summary}
The HERMES DVCS data on beam-charge and beam-spin asymmetries 
is already able to distinguish between some GPD models.
Based on a certain model, the DVCS measurements on transversely polarized
hydrogen lead to a first model dependent constraint for the total angular
momentum of quarks in the nucleon. 
This method, together with increased statistics and improved models should allow
for a constraint with reasonable statistical and theoretical uncertainties in the
future.

\end{document}